\documentclass[pre,preprint]{revtex4}%
\usepackage{amsfonts}
\usepackage{amsmath}
\usepackage{amssymb}
\usepackage{graphicx}%
\begin{document}

\title[Properties of Fermion Systems]{Quasiclassical and Statistical
Properties \\of Fermion Systems}

\author{M. Grigorescu and W. E. Baylis}

\affiliation{Department of Physics, University of Windsor,
Windsor, ON, Canada N9B 3P4}

\keywords{fermion systems, coherent states, Grassmann algebra,
high-T superconductors, tunneling, entangled states}

\pacs{03.65.Sq, 03.65.Ud, 03.65.Yz, 05.30.Fk, 74.72.Dn}

\begin{abstract}
A quasiclassical correspondent for the fermion degrees of freedom is
obtained
by using a time-dependent variational principle with Grassmann coherent
states
as trial functions. In the real parametrization provided by the canonical
coordinates, these states satisfy a closure relationship, and this is used
to
calculate the partition function at finite temperature. The particular
example
considered here consists of a many-fermion system in a quantum
double-well.
Entanglement between the one-particle orbital states leads to deviations
from
the normal Fermi-Dirac distribution. This deviation is reflected in the
dependence of the chemical potential on concentration. In the physical
example
of two adjacent CuO planes in a high-$T_{c}$ superconductor, the
dependence is
suggestive of the pseudogap observed at temperatures $T>T_{c}~.$

\end{abstract}
\volumeyear{2002}
\received[Received text]{date}

\maketitle

\section{Introduction}

The classical limit of an elementary quantum dynamical system can show
spectacular variation in the nature and number of the degrees of freedom.
While the quantum dynamics is mainly stationary, or in transitory regimes
given by a linear superposition of an infinite number of
harmonic-oscillator
modes (the stationary states), the classical dynamics is in general
nonstationary and nonlinear, and it involves a relatively small number of
degrees of freedom.

The transition from quantum to classical statistics in a many-particle
system
at finite temperature, resulting from a change in the underlying
microscopic
dynamics from quantum to classical, is less well explored. The usual
approach
is based on the Wigner-Kirkwood expansion of the quantum partition
function in
powers of the Planck constant $h$. However, convergence of this series is
assured only when the thermal energy $k_{B}T$ is large with respect to the
level spacing \cite{kirk}.

Of course, the classical Maxwell-Boltzmann statistics can be recovered
from
the quantum Bose-Einstein or Fermi-Dirac statistics in the limit of high
temperatures. However, this limit is not exactly a classical limit because
the
constituent particles remain quantum objects. Moreover for
spin-$\frac{1}{2}$
particles, a classical correspondent exists only in the sense of a
pseudomechanics, defined in terms of Grassmann variables \cite{mar}.

Classical degrees of freedom at the quantum level can be introduced with
the
concepts of a molecular (or nuclear) mean field or of spontaneous symmetry
breaking. In such situations, the infinite degeneracy of a metastable
ground
state is lifted by the emergence of a small number of collective
(mean-field)
degrees of freedom. At low energy, the collective motion is quantized, but
with increasing energy, a transition to classical behavior may occur.

The thermodynamics of quantum many-body systems can be affected not only
by
the classical behavior of certain degrees of freedom of the constituent
particles, but also by macroscopic quantum coherence phenomena. Such
phenomena
occur, for example, when a quantum system is split by a potential barrier
into
two entangled components localized in distinct classical regions. This
situation appears in the generic case of two quasi-2D (two-dimensional)
layers
of electron gas separated by a thin insulator. Examples of such bilayer
structures that are particularly important for applications are provided
by
the CuO planes of high-$T_{c}$ superconductors \cite{htc1} and by the
tunnel
junction of the recently developed 2D-2D transistor \cite{tt1,tt2}.

A suitable framework for understanding the quasiclassical behavior of
quantum
systems is provided by the time-dependent variational principle
\cite{kramer}.
This principle leads to quasiclassical dynamics, obtained by constraining
the
quantum dynamics from the infinite-dimensional Hilbert space to a
finite-dimensional manifold of trial functions. The usual mean-field
approximations of many-body theory can be recovered in this framework by
choosing trial functions that are coherent states of a semisimple Lie
group.
However, by the choice of the generators, these approximations provide the
classical analog for particle-hole or particle-particle operators rather
than
for individual fermion fields.

A direct quasiclassical correspondent of the Fermi operators can be
obtained
by using as trial functions the Grassmann coherent states \cite{mar,schw}
. In
Sect. II it will be shown that by a suitable definition of Hermitian
conjugation and of the product between conjugate variables, the Grassmann
coherent states admit a real parameterization. With respect to this
parametrization, the variational principle associates a classical
Hamiltonian
system for each fermion degree of freedom. The integral of the projector
on
the Grassmann coherent states, taken over the invariant volume element
associated with the canonical variables, provides a resolution of the
identity
in the many-fermion Hilbert space (see Appendix). This important
relationship
was obtained previously \cite{klau,mit} by direct integration over
Grassmann
variables. However, the approach based on the Berezin integral \cite{ber}
may
lead to ambiguities, especially in the integration of multiple products
involving dual Grassmann variables. As we show, the present definition
removes
the difficulties encountered in the semiclassical quantization of the
phase
integrals \cite{casa,abe} and leads to the correct Bohr-Wilson-Sommerfeld
formula.

The resolution of the identity allows us to reduce the calculus of the
trace
in many-fermion Hilbert space to an integral over Grassmann coherent
states,
and this is particularly suited for statistical calculations. At the end
of
Sect. II, we show that the present approach leads to the standard
Fermi-Dirac
distribution function.

The effect of quantum entanglement on the particle distribution functions
in
coupled Fermi systems is discussed in Sect. III. The particular coupling
selected here is tunneling through a potential barrier, but this is only a
heuristic model. Similar results can be obtained by using any interaction
that
splits the degenerate levels. It is seen that the level splitting and the
difference between the chemical potentials of the two components combine
in a
single quantity that determines the entangled distribution function. In
the
physical example of the CuO planes in the high-$T_{c}$ compound La$_{2-x}%
$Sr$_{x}$CuO$_{4}$, we show in Sect. IV that the entanglement brings the
normal Fermi-Dirac distribution closer to one of a superconductor. An
observable directly related to the distribution function is the shift of
the
chemical potential as a function of concentration. It is shown that
despite
the simplicity of the model, the shift calculated for an entangled
Fermi-Dirac
distribution is in reasonable agreement with available experimental data.
A
summary of the main results, and the concluding remarks are presented in
Sect. V.

\section{Grassmann coherent states and partition functions}

Consider a model space containing a set of $L$ orthonormal one-particle
fermion states $\left\{  \psi_{i},\ i=1,\ldots,L\right\}  $, and let
$a_{i}^{\dagger}$ ($a_{i}=\left(  a_{i}^{\dagger}\right)  ^{\dagger}$) be
the
associated fermion creation (annihilation) operators in Fock space with
respect to the vacuum state $\left\vert 0\right\rangle $, $a_{i}\left\vert
0\right\rangle =0$, such that $\psi_{i}(x)=\left\langle x\right\vert
a_{i}^{\dagger}\left\vert 0\right\rangle $, $\left\{  a_{i},a_{j}\right\}
=a_{i}a_{j}+a_{j}a_{i}=0$ and $\left\{  a_{i}^{\dagger},a_{j}\right\}
=\delta_{ij}$. Coherent fermion states $\left\vert \zeta\right\rangle $
can be
defined as eigenstates of the fermion annihilation operator $a_{i}$, with
eigenvalues $\zeta_{i}$ \cite{mar,schw}
\begin{equation}
a_{i}\left\vert \zeta\right\rangle =\zeta_{i}\left\vert \zeta\right\rangle
~.
\label{coher}%
\end{equation}
This definition, together with the anticommutivity of the fermion
operators,
leads to
\begin{equation}
\left\{  a_{i},a_{j}\right\}  \left\vert \zeta\right\rangle =\pm\left\{
\zeta_{j},\zeta_{i}\right\}  \left\vert \zeta\right\rangle =0~,
\end{equation}
with the upper ($+$) sign on the right-hand side if $a_{i}$ and
$\zeta_{j}$
commute, $\left[  a_{i},\zeta_{j}\right]  =0$, and the lower ($-$) sign if
they anticommute, $\left\{  a_{i},\zeta_{j}\right\}  =0$. In both cases
$\left\{  \zeta_{i},\zeta_{j}\right\}  =0$, which means that the
\textquotedblleft eigenvalues\textquotedblright\ $\zeta_{i}$,
$i=1,\ldots,L$
cannot be scalars but should be considered Grassmann variables. Thus,
coherent
fermion states exist only in a Grassmannian extension of Fock space.

Let $N$ be the number of independent variables among $\zeta_{i}$,
$i=1,\ldots,L$, and let $\left\{  w_{\alpha}\right\}  $, $\alpha
=1,\ldots,N\leq L$, be a basis in the complex linear space $\Gamma_{N}$ of
the
$\zeta_{i}$. The product of $k\leq N$ distinct basis elements is an
element of
an $(_{k}^{N})$-dimensional linear space $\mathcal{G}_{N}^{k}$. For
example,
$\mathcal{G}_{N}^{0}$ denotes the space of complex scalars and
$\mathcal{G}%
_{N}^{1}\equiv\Gamma_{N}$~. The direct sum
$\bigoplus_{k=0}^{N}\mathcal{G}%
_{N}^{k}$ is the $2^{N}$-dimensional linear space of the full Grassmann
algebra $\mathcal{G}_{N}$.

To calculate matrix elements of many-body fermion operators between
Grassmann
coherent states, we must define the Hermitian conjugation operation
\textquotedblleft$\dagger$\textquotedblright\ not only for operators but
also
for Grassmann variables. The operation is introduced here by using the
linear
space $\Gamma_{N}^{\ast}$ dual to $\Gamma_{N}$, that is the linear space
of
complex linear functionals on $\Gamma_{N}$. The basis in
$\Gamma_{N}^{\ast}$
dual to the basis $\left\{  w_{\alpha}\right\}  $ of $\Gamma_{N}$
comprises
functionals $w_{\alpha}^{\ast}$ defined in terms of unique scalar-valued
mappings \textquotedblleft$\cdot$\textquotedblright\ such that $w_{\alpha
}^{\ast}\cdot w_{\beta}\equiv w_{\beta}\cdot w_{\alpha}^{\ast}=\delta
_{\alpha\beta}$ \cite{doran}. This relation allows us to introduce the
involution operator $\ast:\Gamma_{N}\rightarrow\Gamma_{N}^{\ast}$, related
to
the Hermitian conjugation by $(\zeta_{j}a_{i})^{\dagger}=a_{i}^{\dagger}%
\zeta_{j}^{\ast}$, such that for any $\zeta_{i},\zeta_{j}\in\Gamma_{N}$,
$(\zeta_{j}^{\ast})^{\ast}=\zeta_{j}$, $(\zeta_{i}\zeta_{j})^{\ast}=\zeta
_{j}^{\ast}\zeta_{i}^{\ast}$, and $(z\zeta_{j})^{\ast}=z^{\ast}\zeta_{j}%
^{\ast}$, where $z$ is any complex number and $z^{\ast}$ is its complex
conjugate.

It is often convenient to think of the scalar-valued mappings denoted by
\textquotedblleft$\cdot$\textquotedblright\ as products in the direct-sum
space $\mathcal{C}_{N}=\Gamma_{N}^{\ast}\oplus\Gamma_{N}.$ Duality then
assigns a complex scalar value to any \emph{product} $\zeta_{j}^{\ast}%
\cdot\zeta_{k}=\sum_{\alpha=1}^{N}z_{j}^{\alpha\ast}z_{k}^{\alpha}$ of the
Grassmann variables $\zeta_{j}^{\ast}=\sum_{\alpha=1}^{N}z_{j}^{\alpha\ast
}w_{\alpha}^{\ast}\in\Gamma_{N}^{\ast}$ and $\zeta_{k}=\sum_{\alpha=1}%
^{N}z_{k}^{\alpha}w_{\alpha}\in\Gamma_{N}.$ The space $\mathcal{C}_{N}$
generates a $2^{2N}$-dimensional Clifford algebra $C\!\ell_{N,N}$
\cite{doran,bay,loun} in which the scalar product of \emph{null vectors}
$\zeta_{j}^{\ast}$ and $\zeta_{k}$ can be identified with the
anticommutator:
$\zeta^{\ast}\cdot\zeta^{\prime}=\frac{1}{2}\left\{  \zeta^{\ast}%
,\zeta^{\prime}\right\}  $ of their associative Clifford product. Physical
quantities are evaluated as the scalar part of expressions in
$C\!\ell_{N,N}$,
and unless otherwise noted, this restriction is assumed in expressions of
expectation values below.

When $L=1$, an unnormalized Grassmann coherent state can be expressed as%
\begin{equation}
\left\vert \tilde{\zeta}\right\rangle =e^{\pm\zeta_{1}a_{1}^{\dagger}%
}\left\vert 0\right\rangle =\left(  1\pm\zeta_{1}a_{1}^{\dagger}\right)
\left\vert 0\right\rangle =\left\vert 0\right\rangle
\pm\zeta_{1}\left\vert
1\right\rangle , \label{ungcs}%
\end{equation}
where $\left\vert 1\right\rangle =a_{1}^{\dag}\left\vert 0\right\rangle .$
Since the (scalar part of the) overlap
\begin{equation}
\left\langle \tilde{\zeta}|\tilde{\zeta}\right\rangle =1+\zeta_{1}^{\ast}%
\cdot\zeta_{1}%
\end{equation}
is a real number, the normalized state (\ref{coher}) is
\begin{equation}
\left\vert \zeta\right\rangle =\frac{\left(  1\pm\zeta_{1}a_{1}^{\dagger
}\right)  \left\vert 0\right\rangle
}{\sqrt{1+\zeta_{1}^{\ast}\cdot\zeta_{1}}%
}.
\end{equation}
It may be verified that since $a_{1}a_{1}^{\dagger}\left\vert
0\right\rangle
=\left\{  a_{1},a_{1}^{\dagger}\right\}  \left\vert 0\right\rangle
=\left\vert
0\right\rangle $ and $\zeta_{1}^{2}=0,$%
\begin{align}
a_{1}\left\vert \zeta\right\rangle  &  =\frac{\zeta_{1}a_{1}a_{1}^{\dagger
}\left\vert 0\right\rangle }{\sqrt{1+\zeta_{1}^{\ast}\cdot\zeta_{1}}}%
=\frac{\zeta_{1}\left\vert 0\right\rangle }{\sqrt{1+\zeta_{1}^{\ast}\cdot
\zeta_{1}}}\nonumber\\
&  =\frac{\zeta_{1}\left(  1\pm\zeta_{1}a_{1}^{\dagger}\right)  \left\vert
0\right\rangle
}{\sqrt{1+\zeta_{1}^{\ast}\cdot\zeta_{1}}}=\zeta_{1}\left\vert
\zeta\right\rangle \,.
\end{align}
In the $L$-level case it is convenient to assume that $\zeta_{i}$ and
$a_{j}^{\dagger}$ anti-commute, and we therefore choose the
\textquotedblleft%
$-$\textquotedblright\ sign in Eq. (\ref{ungcs}). Then the normalized
Grassmann coherent state has the general expression
\begin{equation}
\left\vert \zeta\right\rangle =\prod_{k=1}^{L}\frac{\left(  1-\zeta_{k}%
a_{k}^{\dagger}\right)
}{\sqrt{1+\zeta_{k}^{\ast}\cdot\zeta_{k}}}\left\vert
0\right\rangle =\prod_{k=1}^{L}\frac{\exp\left(  -\zeta_{k}a_{k}^{\dagger
}\right)  }{\sqrt{1+\zeta_{k}^{\ast}\cdot\zeta_{k}}}\left\vert
0\right\rangle
. \label{zeta}%
\end{equation}
The expectation value of the particle number operator
$a_{k}^{\dagger}a_{k}$
in this state is
\begin{equation}
n_{k}=\left\langle \zeta\right\vert a_{k}^{\dagger}a_{k}\left\vert
\zeta\right\rangle =\left\langle \zeta\right\vert \zeta_{k}^{\ast}\zeta
_{k}\left\vert \zeta\right\rangle =\frac{\zeta_{k}^{\ast}\cdot\zeta_{k}%
}{1+\zeta_{k}^{\ast}\cdot\zeta_{k}}\,.
\end{equation}
Note that $\zeta_{k}^{\ast}\zeta_{k}$ is not a pure scalar in
$C\!\ell_{N,N}$
(it also contains a bivector part), and since the coherent state
$\left\vert
\zeta\right\rangle $ is also not a Grassmann scalar, we do not restrict
$\zeta_{k}^{\ast}\zeta_{k},$ but only its expectation value $\left\langle
\zeta\right\vert \zeta_{k}^{\ast}\zeta_{k}\left\vert \zeta\right\rangle ,$
to
its scalar part. When $N=L$, it is possible to choose
$\zeta_{k}=z_{k}w_{k}$.
In this case, $\zeta_{k}^{\ast}\cdot\zeta_{k}=z_{k}^{\ast}z_{k}$ and the
absolute value of the complex variable $z_{k}=\rho_{k}e^{i\phi_{k}}$ is
related to the particle number by $n_{k}=\rho_{k}^{2}/\left(  1+\rho_{k}%
^{2}\right)  $. This relation limits $n_{k}$ to the range from $0$ (when
$\rho_{k}=0$ ) to $1$ (when $\rho_{k}\rightarrow\infty$ ). With such
parametrization, Grassmann coherent states become $su\left(  2\right)  $
coherent states (see Appendix). The values of $n_{k}$ are quantized by the
anticommutation relation $\left\{  a_{i}^{\dagger},a_{j}\right\}  =\delta
_{ij}$ or by the old quantum theory (see below).

A quasiclassical approximation of the quantum many-fermion dynamics can be
obtained by using the Grassmann coherent state $\left\vert
\zeta\right\rangle
$ as trial function in the time-dependent variational principle. If
$\left\vert \zeta\right\rangle $ depends on time only through the complex
scalar variables $z_{k}$, then the best approximation for an exact
solution of
the time-dependent Schr\"{o}dinger equation satisfies
\begin{equation}
\delta_{x}\int_{t_{1}}^{t_{2}}\left\langle \zeta\right\vert i\hbar\partial
_{t}-H\left\vert \zeta\right\rangle dt=0~, \label{variation}%
\end{equation}
where $H$ is the Hamiltonian of the many-fermion system, $t_{1},t_{2}$ are
fixed times, and $\delta_{x}$ denotes the variation of the integral with
respect to the real functions of time $\rho_{k}(t)$ and $\phi_{k}(t)$,
$k=1,\ldots,L$, denoted together by $x_{k},\ k=1,\ldots,2L$.

The equations of motion determined by Eq. (\ref{variation}) have the
general
form
\begin{equation}
\sum_{i=1}^{2L}\dot{x}_{i}\omega_{ij}=\frac{\partial\mathcal{H}}{\partial
x_{j}}~,
\end{equation}
where $\mathcal{H}=\left\langle \zeta\right\vert H\left\vert \zeta
\right\rangle $, and $\omega_{ij}=-\omega_{ji}=2\hbar\Im(\left\langle
\partial_{i}\zeta\right.  \left\vert \partial_{j}\zeta\right\rangle )$ are
the
coefficients of a nondegenerate 2-form (a symplectic structure) on the
manifold of Grassmann coherent states. A canonical transformation to the
new
variables
\begin{equation}
q_{k}=\sqrt{\frac{2\hbar}{1+\rho_{k}^{2}}}\sin\phi_{k}~,\;p_{k}=\sqrt
{\frac{2\hbar}{1+\rho_{k}^{2}}}\cos\phi_{k}~, \label{pkqk}%
\end{equation}
defined on the disk of radius $\sqrt{2\hbar}$ in the $pq$ plane, reduces
further the equations of motion to
\begin{equation}
\dot{q}_{k}=\frac{\partial\mathcal{H}}{\partial p_{k}}~,\;\dot{p}_{k}%
=-\frac{\partial\mathcal{H}}{\partial q_{k}}~.
\end{equation}
If the states $\psi_{i}$, $i=1,\ldots,L$ are one-particle energy
eigenstates
with eigenvalues $\epsilon_{i}$, then a many-fermion Hamiltonian of the
form
$H_{f}=\sum_{k=1}^{L}\epsilon_{k}a_{k}^{\dagger}a_{k}$ can be considered
as
the Hamiltonian of a \textquotedblleft Fermi
oscillator.\textquotedblright%
\ Since%
\begin{equation}
q_{k}^{2}+p_{k}^{2}=\frac{2\hbar}{1+\rho_{k}^{2}}=2\hbar\left(  1-n_{k}%
\right)  ,
\end{equation}
for the Fermi oscillator $\mathcal{H}_{f}=\sum_{k=1}^{L}\epsilon_{k}\left[
1-\left(  p_{k}^{2}+q_{k}^{2}\right)  /2\hbar\right]  $, and the equations
of
motion have the solution $\rho_{k}\left(  t\right)  =\mathrm{const}$,
$\phi_{k}\left(  t\right)  =-t\epsilon_{k}/\hbar$. This represents a
harmonic
oscillation of the variables $p_{k}$ and $q_{k}$ at the angular frequency
$\epsilon_{k}/\hbar$. Moreover, in this particular case, the Grassmann
coherent state $\left\vert \zeta\right\rangle \left(  t\right)  $ is not
an
approximation but an exact solution of the time-dependent Schr\"{o}dinger
equation,
\begin{equation}
i\hbar\frac{\partial\left\vert \zeta\right\rangle }{\partial t}=H_{f}%
\left\vert \zeta\right\rangle .
\end{equation}

The \textquotedblleft classical\textquotedblright\ variables $p_{k}$ and
$q_{k}$, have been introduced without reference to the physical degrees of
freedom of the particles (space coordinates + spin). However, it is
interesting to note that they behave in all respects as real classical
variables. In the $pq$ plane the orbits are circles of radius $\sqrt
{2\hbar\left(  1-n_{k}\right)  }$. The Bohr-Wilson-Sommerfeld quantization
formula selects only the orbits having an area which is an integer
multiple of
the Planck constant $h$, such that
\begin{equation}
2\pi\hbar\left(  1-n_{k}\right)  =nh~,\;n=0,1,....
\end{equation}
This shows that $1-n_{k}$ should be a non-negative integer, and the only
possible values of $n_{k}$ are $0$ and $1$.

The remarkable property of Grassmann coherent states of providing a
resolution
of the identity operator in many-body Hilbert space (see Appendix)
\begin{equation}
\mathcal{I}=\frac{1}{\pi^{L}}\int d^{L}\Omega\left\vert \zeta\right\rangle
\left\langle \zeta\right\vert
\end{equation}
makes them particularly suitable for applications in statistical
mechanics.
The partition function $Z$ of a quantum $N_{p}$-body system is defined by
the
trace in Fock space, $Z=Tr(e^{-\beta H})=\sum_{\left[  n\right]
}\left\langle
\left[  n\right]  \right\vert e^{-\beta H}\left\vert \left[  n\right]
\right\rangle ,$ where $\beta=1/k_{B}T$, $H$ is the many-body Hamiltonian
operator, and $\left\vert \left[  n\right]  \right\rangle \equiv\left\vert
n_{1}n_{2}...\right\rangle $ represents one of a complete set of states,
labeled by the occupation numbers $n_{k}$ of the single-particle levels,
and
restricted by the condition $\sum_{k}n_{k}=N_{p}$~. If these states are
eigenstates of $H$ with eigenvalues $E_{[n]}$, then $Z=\sum_{[n]}e^{-\beta
E_{[n]}}$. Moreover, if $E_{[n]}=\sum_{k}n_{k}\epsilon_{k}$, then
$Z=\sum_{\left[  n\right]  }\prod_{k}e^{-\beta n_{k}\epsilon_{k}}$. The
sum
$\Sigma_{\lbrack n]}$ can be calculated if the condition of having a fixed
number of particles is relaxed and the $n_{k}$ are independent
\cite{somm}.
The average number of particles can still be fixed by introducing the
chemical
potential $\mu$; then $Z=\prod_{k}Z_{k}$ becomes the product of one-level
partition functions $Z_{k}=\Sigma_{n_{k}}\exp\left[  -\beta n_{k}\left(
\epsilon_{k}-\mu\right)  \right]  $ with $n_{k}=0,1$ for Fermi-Dirac
statistics and $n_{k}=0,...,\infty$ for Bose-Einstein.

The statistical significance of the \textquotedblleft Fermi
oscillator\textquotedblright\ appears when the corresponding partition
function $Z_{f}$ is calculated as an integral over the manifold of
Grassmann
coherent states. The resolution of the identity presented in the Appendix
allows us to write
\begin{align}
Z_{f}  &  =\mathrm{Tr}\left[  e^{-\beta\left(  H_{f}-\mu N_{T}\right)
}\right]  =\prod_{k=1}^{L}Z_{k}\\
&  =\frac{1}{\pi^{L}}\int d^{L}\Omega\left\langle \zeta\right\vert
e^{-\beta\left(  H_{f}-\mu N_{T}\right)  }\left\vert \zeta\right\rangle
\,,\nonumber
\end{align}
where $N_{T}=\sum_{k}a_{k}^{\dag}a_{k}$ is the total particle number
operator
and
\begin{equation}
Z_{k}=\frac{1}{\pi}\int d\Omega_{k}\left\langle \zeta_{k}\right\vert
e^{-\beta
a_{k}^{\dagger}a_{k}\left(  \epsilon_{k}-\mu\right)  }\left\vert \zeta
_{k}\right\rangle \,.
\end{equation}
Because $d\Omega_{k}=d\xi_{k}d\phi_{k}=dp_{k}dq_{k}/\hbar$, and for any
integer power $n$, $\left\langle \zeta_{k}\right\vert \left(
a_{k}^{\dagger
}a_{k}\right)  ^{n}\left\vert \zeta_{k}\right\rangle =\left\langle \zeta
_{k}\right\vert a_{k}^{\dagger}a_{k}\left\vert \zeta_{k}\right\rangle
=n_{k}$
with $n_{k}=1-\left(  p_{k}^{2}+q_{k}^{2}\right)  /2\hbar$, we get
\begin{align}
Z_{k}  &  =\frac{1}{\pi\hbar}\int_{p_{k}^{2}+q_{k}^{2}\leq2\hbar}dp_{k}%
dq_{k}\left\{  1+\left[  e^{-\beta\left(  \epsilon_{k}-\mu\right)
}-1\right]
\left(  1-\frac{p_{k}^{2}+q_{k}^{2}}{2\hbar}\right)  \right\} \nonumber\\
&  =1+e^{-\beta\left(  \epsilon_{k}-\mu\right)  }%
\end{align}
recovering the standard result for fermions.

If $\epsilon_{k}-\mu$, $k=1,\ldots,L,$ and $T$ are such that $\left\vert
\epsilon_{k}-\mu\right\vert <<k_{B}T$, then $e^{-\beta\left(
\epsilon_{k}%
-\mu\right)  }-1\approx-\beta\left(  \epsilon_{k}-\mu\right)  $. In this
case,
$Z_{f}$ takes the form of a classical partition function, defined by the
phase-space integral of $\exp\left[  -\beta\left(  \mathcal{H}_{f}%
-\mu\mathcal{N}\right)  \right]  $ with respect to the volume element
$d^{L}\Omega$, where $\mathcal{N}$ is the expectation value of $N_{T}\,.$

\section{Distribution functions in entangled Fermi systems}

The interplay among classical localization, quantum coherence, and
temperature
on the one-particle distribution function plays a central role in the
many-fermion fusion or fission reactions and in the operation of tunnel
junctions. Localization occurs when a container filled with Fermi gas is
split
into two identical, classically distinct components. The opposite process,
when two separated, identical Fermi systems merge to form a single large
system, represents delocalization. In both situations, the intermediate
stages
can be modeled by two entangled Fermi systems confined in classically
distinct
potential wells that are separated by a finite barrier.

The problem of a system of fermions placed in a double-well potential at
zero
temperature was studied previously in the context of the predicted
exchange-driven spontaneous transitions from bilayer to monolayer
ground-state
configurations \cite{zh}. The influence of coupling on the distribution
function at finite temperature is studied here by considering a system of
fermions of mass $M$ placed in a one-dimensional potential, consisting of
two
identical rectangular potential wells of width $w$ and depth $U,$
separated by
a distance $d$. The bounded one-particle energy levels in this case are
$E_{k}=x_{k}^{2}U,$ where $x=x_{k}$ satisfies
\begin{equation}
\tan\left(  \tilde{w}x\right)  =\frac{2x\sqrt{1-x^{2}}}{\pm e^{-\tilde{d}%
\sqrt{1-x^{2}}}+2x^{2}-1}%
\end{equation}
with $\tilde{w}=w\sqrt{2MU/\hbar^{2}}$ and
$\tilde{d}=d\sqrt{2MU/\hbar^{2}}$.
For the fully merged system ($d=0$, $\tilde{w}_{d}=2\tilde{w}$) as well as
for
each single well ($d=\infty$, $\tilde{w}_{d}=\tilde{w}$), the $x_{k}$ are
solutions of%
\begin{equation}
\tan\left(  \tilde{w_{d}}x\right)  =\frac{2x\sqrt{1-x^{2}}}{2x^{2}-1}.
\end{equation}
Let us denote by $\left\vert k1\right\rangle $ and $\left\vert
k2\right\rangle
$ the orbital wave functions localized in wells 1 and 2, respectively,
which
at $d=\infty$ correspond to the same energy level $\epsilon_{k}^{0}=x_{k}%
^{2}U$. At large but finite $d$, tunneling lifts the degeneracy. The
eigenstates of the coupled system are represented by the entangled
symmetric
and antisymmetric superpositions $\left\vert k_{\pm}\right\rangle
=(\left\vert
k1\right\rangle \pm\left\vert k2\right\rangle )/\sqrt{2}$, separated in
energy
by the tunnel splitting $\Delta_{k}=\epsilon_{k}^{-}-\epsilon_{k}^{+}%
\approx2V_{c}(k)$, where
\begin{equation}
V_{c}(k)=\frac{x_{k}U}{\tilde{w}}e^{-\tilde{d}\sqrt{1-x_{k}^{2}}%
}\label{tunnel}%
\end{equation}
is the tunneling matrix element. The Hamiltonian of a many-fermion system
distributed over the levels of the entangled wells is
$H_{e}=\sum_{km}h_{km}$,
where $m=\pm1/2$ is the $z$-component of the spin, and
\begin{align}
h_{km} &  =\epsilon_{k}^{-}a_{k_{-}m}^{\dagger}a_{k_{-}m}+\epsilon_{k}%
^{+}a_{k_{+}m}^{\dagger}a_{k_{+}m}\\
&  =\epsilon_{k}^{0}\left(  a_{1km}^{\dagger}a_{1km}+a_{2km}^{\dagger}%
a_{2km}\right)  +\frac{\Delta_{k}}{2}\left(  a_{1km}^{\dagger}a_{2km}%
+a_{2km}^{\dagger}a_{1km}\right)  .\nonumber
\end{align}
In a quantum description of the entangled system, the corresponding
partition
function is $Z_{e}=\mathrm{Tr}(e^{-\beta(H_{e}-\mu N_{T})})$. However, as
long
as the two wells remain classically distinct near equilibrium, we can
assume
that each one develops its own chemical potential such that $Z_{e}%
=\mathrm{Tr}(e^{-\beta(H_{e}-\mu_{1}N_{1}-\mu_{2}N_{2})})$. Following the
procedure presented in Sect. II, $Z_{e}$ can be reduced to an integral
over
Grassmann coherent states,
\begin{align}
Z_{e} &  =\frac{1}{\pi^{2L}}\int
d^{L}\Omega_{1}d^{L}\Omega_{2}\left\langle
\zeta\right\vert e^{-\beta(H_{e}-\mu_{1}N_{1}-\mu_{2}N_{2})}\left\vert
\zeta\right\rangle \nonumber\\
&  =\prod_{k=1}^{L}Z_{k\frac{1}{2}}Z_{k-\frac{1}{2}}~,
\end{align}
where
\begin{equation}
Z_{km}=\frac{1}{\pi^{2}}\int d\Omega_{1km}d\Omega_{2km}\left\langle \zeta
_{km}\right\vert e^{-\beta(h_{km}-\mu_{1}a_{1km}^{\dagger}a_{1km}-\mu
_{2}a_{2km}^{\dagger}a_{2km})}\left\vert \zeta_{km}\right\rangle
\end{equation}
and
\begin{equation}
\left\vert \zeta_{km}\right\rangle =\frac{\exp\left(  -\zeta_{1km}%
a_{1km}^{\dagger}-\zeta_{2km}a_{2km}^{\dagger}\right)  }{\sqrt{(1+\zeta
_{1km}^{\ast}\cdot\zeta_{1km})(1+\zeta_{2km}^{\ast}\cdot\zeta_{2km})}%
}\left\vert 0\right\rangle \,.
\end{equation}
The operator $\exp\left[  -\beta\left(  h_{km}-\mu_{1}a_{1km}^{\dagger}%
a_{1km}-\mu_{2}a_{2km}^{\dagger}a_{2km}\right)  \right]  $ has the
expansion
\begin{align}
&  1+(e^{\alpha_{k}}\cosh\eta_{k}-1)(a_{1km}^{\dagger}a_{1km}+a_{2km}%
^{\dagger}a_{2km})\nonumber\\
&  +e^{\alpha_{k}}\sinh\eta_{k}[\frac{r}{\eta_{k}}(a_{1k}^{\dagger}%
a_{1k}-a_{2k}^{\dagger}a_{2k})\nonumber\\
&  +\frac{s_{k}}{\eta_{k}}(a_{1km}^{\dagger}a_{2km}+a_{2km}^{\dagger}%
a_{1km})]\\
&  +(e^{2\alpha_{k}}-2e^{\alpha_{k}}\cosh\eta_{k}+1)a_{1km}^{\dagger}%
a_{1km}a_{2km}^{\dagger}a_{2km}\,,\nonumber
\end{align}
where $\alpha_{k}=-\beta(\epsilon_{k}^{0}-(\mu_{1}+\mu_{2})/2)$, $\eta
_{k}=\sqrt{s_{k}^{2}+r^{2}},$ $s_{k}=\beta\Delta_{k}/2$, and $r=\beta(\mu
_{1}-\mu_{2})/2$. Denoting by $\left\langle \left\langle A\right\rangle
\right\rangle $ the integral
\begin{equation}
\left\langle \left\langle A\right\rangle \right\rangle
=\frac{1}{\pi^{2}}\int
d\Omega_{1km}d\Omega_{2km}\left\langle \zeta_{km}\left\vert A\right\vert
\zeta_{km}\right\rangle
\end{equation}
we get
\begin{equation}
\left\langle \left\langle 1\right\rangle \right\rangle =4~,\;\left\langle
\left\langle a_{1km}^{\dagger}a_{1km}\right\rangle \right\rangle
=\left\langle
\left\langle a_{2km}^{\dagger}a_{2km}\right\rangle \right\rangle =2~,
\end{equation}%
\[
\left\langle \left\langle a_{1km}^{\dagger}a_{1km}a_{2km}^{\dagger}%
a_{2km}\right\rangle \right\rangle =1
\]
such that $Z_{km}$ has the expression
\begin{equation}
Z_{km}=e^{2\alpha_{k}}+2e^{\alpha_{k}}\cosh\eta_{k}+1~.\label{Zkm}%
\end{equation}
The quantity that contains the effect of entanglement on the distribution
function is $\eta_{k}=\sqrt{s_{k}^{2}+r^{2}}$. If $|s_{k}|<<|r|$, then
$Z_{km}$ becomes the product $Z_{k1m}Z_{k2m}$ of the Fermi distributions
for
the orbital levels $\left\vert k1\right\rangle $ and $\left\vert
k2\right\rangle $,
\begin{align}
Z_{k1m} &  =1+e^{-\beta(\epsilon_{k}^{0}-\mu_{1})}\\
Z_{k2m} &  =1+e^{-\beta(\epsilon_{k}^{0}-\mu_{2})}.\nonumber
\end{align}
When $|r|<<|s_{k}|$, then $Z_{km}$ is the product $Z_{k_{-}m}Z_{k_{+}m}$
of
the distributions
\begin{align}
Z_{k_{-}m} &  =1+e^{-\beta(\epsilon_{k}^{-}-\mu)}\\
Z_{k_{+}m} &  =1+e^{-\beta(\epsilon_{k}^{+}-\mu)}\nonumber
\end{align}
and $\mu=(\mu_{1}+\mu_{2})/2$ is the Fermi level of the fully merged
system.

For the electron layers in the 2D-2D tunneling transistor, the difference
between the chemical potentials is related to the applied voltage $V$,
$\mu_{1}-\mu_{2}=eV$ \cite{tt2}. In the case of two adjacent CuO layers of
a
high-$T_{c}$ compound, a small difference may appear due to the slight
asymmetry observed in the placement of the oxygen acceptors between the
planes
\cite{kastner}.

\section{Entangled Fermi distributions in high-$T_{c}$ superconductors}

In high-$T_{c}$ materials the charge carriers are confined to the CuO
planes,
but the interlayer coupling is important in both the normal and
superconducting states. At intermediate temperatures and low doping, the
interlayer coupling is dominated by the magnetic interaction of the
electron
spins, producing antiferromagnetic ordering. The charge transport between
the
layers, practically absent in this state, is assumed to have both a
coherent,
momentum-conserving component and an incoherent part due to
impurity-assisted
hopping \cite{kim}. With increasing doping, at lower temperature the
coupling
takes the form of coherent Josephson pair tunneling, supposedly one of the
basic phenomena responsible for the superconductivity at high
temperatures. In
this perspective, the observed decrease of transition temperature $T_{c}$
with
an increase of the number of CuO planes seems to be explained by the loss
of
the phase coherence \cite{pav}.

Layered copper oxide compounds are peculiar not only in their high $T_{c}$
values compared to conventional superconductors, but also in various
anomalous
properties observed in their normal state. Anomalies showing a common
trend
with respect to the temperature and to concentration of charge carriers
are
confirmed by a wide variety of experimental techniques \cite{bs}. Many
anomalous features can be interpreted in terms of a \textquotedblleft
normal-state gap\textquotedblright\ or pseudogap \cite{mit,bs,pwa} of
unknown
origin that turns into the BCS gap when the material becomes
superconducting.

The results presented in Sect. III indicate that entanglement between
adjacent
layers can produce significant deviations from the standard Fermi-Dirac
distribution that may contribute to the observed anomalies. The magnitude
of
such effects is evaluated here within a schematic model in which the
energy of
the electrons in a CuO plane without coupling has the simple form
\begin{equation}
\epsilon_{k}^{0}\equiv\epsilon_{\kappa}^{0}=\epsilon_{p}+\epsilon_{n}^{c}\,,
\end{equation}
where $\kappa\equiv(\epsilon_{p},n)$ denotes the couple of a continuous
index
$\epsilon_{p}$ and a discrete index $n$. These indices correspond to the
terms
$\epsilon_{p}$ and $\epsilon^{c}$ of the total energy due to the motion of
the
electrons along directions parallel and normal to the planes,
respectively.
The in-plane energy term is approximated by a single band, $\epsilon_{p}%
\in\lbrack\epsilon_{min},\epsilon_{max}]$, with $-\epsilon_{min}%
=\epsilon_{max}=1$ eV, and a Fermi level $\epsilon_{F}=-0.45$ eV
\cite{ck}.
Confinement is modeled by a one-dimensional potential well placed normal
to
the planes. As suggested by the expression for the hole binding potential
in
oxygenated La$_{2}$CuO$_{4+y}$, \cite{kastner}, the potential well is
about 35
meV deep and has a radius $a=8$ \AA . Thus, $\epsilon_{k}^{c}$ is
calculated
by using the square-well estimate $\epsilon_{n}^{c}=-(1-x_{n}^{2})U$,
$x_{n}\approx n\pi/\tilde{w}<1$, where $U=35$ meV, and $w=2a=16$ \AA . The
interlayer coupling matrix element $t_{c}$ depends on doping, and in
La$_{2-x}$Sr$_{x}$CuO$_{4}$, $t_{c}\sim0.66x^{2.974}$ eV \cite{cooper}.
This
shows that $t_{c}$ depends on the hole concentration $\delta\approx x$ and
on
the band filling $f=1-\delta$.

The anomalous behavior attributed to the formation of a pseudogap occurs
in
general in underdoped compounds. In La$_{2-x}$Sr$_{x}$CuO$_{4},$ $T_{c}$
attains a maximum of $\sim40$ K at $x_{opt}\sim0.15$. Measurements of the
specific heat coefficient \cite{loram} in this compound indicate the
existence
of a pseudogap for $x<0.22$. In the range $0.1<x<0.22,$ the pseudogap has
an
almost linear decrease with $x$, from $\sim200$ meV to 0 \cite{fuji}. At
$x\sim0.1,$ when the pseudogap is large, the interlayer coupling strength
is
$t_{c}\sim1$ meV. This value is large compared to the one obtained from
Eq.
(\ref{tunnel}), indicating that it may include contributions due to other
interactions, stronger than tunneling coupling. In the present
calculations,
the energy splitting $\Delta_{\kappa}\equiv\Delta_{n}=2V_{c}(n)$ was
calculated by taking $V_{c}(n)=0.001+1.5\sqrt{x_{n}}$ eV. This includes
the
realistic term $t_{c}=1$ meV, and a term weakly increasing with the
excitation
energy along the normal to the plane.

The number of particles $N_{1}$ in the CuO plane 1 is related to the
partition
function $Z_{\kappa m}$ of Eq. (\ref{Zkm}) by
\begin{equation}
N_{1}=\frac{\partial\ln Z}{\partial\beta\mu_{1}}=\sum_{m}\sum_{\kappa}%
\frac{\partial\ln Z_{\kappa m}}{\partial\beta\mu_{1}}~.
\end{equation}
Here $\sum_{\kappa}$ is calculated as an integral over the planar band and
a
sum over the bounded energy levels for motion normal to the plane, such
that
\begin{equation}
\sum_{\kappa}\equiv\sum_{n}N_{p}S\int d\epsilon_{p}~,
\end{equation}
where $N_{p}=m^{\ast}/(2\pi\hbar^{2})$ is the planar density of states,
$m^{\ast}$ is the effective band mass \cite{ino}, and $S$ is the plane
surface. Thus, the planar density $D_{p}=N_{1}/S$ of charge carriers takes
the
form
\begin{equation}
D_{p}=\int_{\epsilon_{min}}^{\epsilon_{max}}d\epsilon\left\langle
n\right\rangle _{(\epsilon)}~,
\end{equation}
where
\begin{equation}
\left\langle n\right\rangle _{(\epsilon)}=2N_{p}\sum_{n}\frac{e^{\alpha
_{\epsilon n}}+\cosh\eta_{n}+r/\eta_{n}\sinh\eta_{n}}{e^{\alpha_{\epsilon
n}%
}+2\cosh\eta_{n}+e^{-\alpha_{\epsilon n}}},
\end{equation}
$\alpha_{\epsilon n}=-\beta(\epsilon+\epsilon_{n}^{c})+\beta(\mu_{1}+\mu
_{2})/2$, $\eta_{n}=\sqrt{s_{n}^{2}+r^{2}}$, $s_{n}=\beta\Delta_{n}/2,$
and
$r=\beta(\mu_{1}-\mu_{2})/2$. The density $\left\langle n\right\rangle
_{(\epsilon)}$ is shown as a function of $\epsilon$ at $T=120$ K and $r=0$
in
Fig. 1 (c). For comparison, on the same plot we display (a) the standard
Fermi-Dirac distribution obtained in the absence of interlayer coupling
($V_{c}=0$, $s_{n}=0$), and (b) the distribution
\begin{equation}
\left\langle n\right\rangle _{(\epsilon)}^{sc}=N_{p}\sum_{n}[1-\frac
{\epsilon_{\kappa}^{0}-\epsilon_{F}}{E_{\kappa}}+\frac{2(\epsilon_{\kappa}%
^{0}-\epsilon_{F})}{E_{\kappa}(e^{\beta E_{\kappa}}+1)}]
\end{equation}
of a superconducting system at finite temperature \cite{blatt}. In this
expression, the quasiparticle energy $E_{\kappa}=\sqrt{(\epsilon_{\kappa}%
^{0}-\epsilon_{F})^{2}+\Delta_{sc}^{2}}$ was calculated by using a
fictitious
superconducting gap $\Delta_{sc}=0.6$ eV.

The comparison between these functions shows that although different, as
far
as deviations from the normal Fermi-Dirac distribution are concerned, (b)
and
(c) may account for similar effects. An observable related directly to the
particle distribution is represented by the shift $\Delta\mu$ of the
chemical
potential as a function of concentration. This shift was subject to a
detailed
experimental investigation, observing that in La$_{2-x}$Sr$_{x}$CuO$_{4}$,
an
anomaly occurs at low doping levels \cite{ino}.

The comparison of the experimental data with the shift determined by the
distribution functions (a),(b) and (c) of Fig. 1 is presented in Fig. 2.
The
reference Fermi level was fixed at $\epsilon_{F}=-0.45$ eV, and the hole
concentration $\delta=1-f$ was calculated by assuming that $f=D_{p}|_{\mu
_{1}=\mu_{2}=\epsilon_{F}+\delta\mu}/D_{p}|_{\mu_{1}=\mu_{2}=\epsilon_{F}}$.
The Fermi-Dirac distribution produces a linear decrease (Fig. 2-a), while
$\left\langle n\right\rangle _{(\epsilon)}^{sc}$ (Fig. 2-b), and
$\left\langle
n\right\rangle _{(\epsilon)}$ (Fig. 2-c), both lead to a dependence
$\Delta
\mu\sim\delta^{2}$. The experimental values can be fitted by a quadratic
polynomial (Fig. 2-d).

The suppression of the chemical potential shift for small $\delta$
compared to
its value at overdoping is taken as an indication for the opening of the
pseudogap \cite{ino}. The present calculations indicate that the shift
produced by an entangled Fermi-Dirac distribution can be very well
simulated
by introducing a fictitious gap parameter. The entangled distribution
leads
also to a quadratic dependence of $\Delta\mu$ on $\delta$, as seen in
experiment, in contrast to the standard Fermi-Dirac distribution. However,
although overall the results are in a reasonable agreement with
experimental
data, the simple model considered here cannot explain the complete
suppression
of the chemical potential shift for small $\delta$.

\section{Summary and Conclusions}

The effective number of microscopic degrees of freedom in a many-particle
system may change when classical and quantum aspects coexist, and such
variations are reflected in the expression of thermodynamic and
statistical
functions. For a classical particle, there are few translational and
rotational degrees of freedom. The classical motion can, in principle, be
described in terms of an infinite number of fictitious harmonic
oscillators in
a Fourier transform, but these oscillators have no statistical relevance.
The
same holds to a certain degree in the case of a quantum object. The energy
eigenstates are normalized, and have an oscillatory time evolution.
However,
as shown by the quasiclassical picture of the old quantum mechanics, these
eigenstate oscillators provide only a partition of the low-dimensional
classical phase space into cells of the size of the Planck constant $h$
rather
than representing new physical degrees of freedom. However, new degrees of
freedom are represented by the occupation numbers of these cells.

The quasiclassical dynamics of the occupation numbers in a system of
fermions
at finite temperature was derived here using Grassmann coherent states. We
have shown that these states admit a real parametrization in which they
represent $su(2)$ coherent states (Appendix). With respect to this
parametrization the expectation value of the occupation number obeys
Hamiltonian dynamics (Sect. II), and the Grassmann coherent states satisfy
a
closure relationship (Appendix). This relationship allows us to recover
the
standard Fermi-Dirac distribution function (Sect. II) and to calculate the
particle density in entangled Fermi systems at finite temperature (Sect.
III).

It is interesting to note that within the thermo-field dynamics formalism
\cite{tfd}, the trace is calculated as a \textquotedblleft thermal
vacuum\textquotedblright\ expectation value. This \textquotedblleft
temperature-dependent vacuum\textquotedblright\ is defined in terms of the
composite system consisting of the original physical system and an
identical
but fictitious quantum dynamical system. When the fermion operators of the
fictitious system are replaced by Grassmann variables, the
\textquotedblleft
thermal vacuum\textquotedblright\ takes the form of a Grassmann coherent
state.

The entangled Fermi-Dirac distribution can be relevant in the description
of
coupled 2D-electron systems. As an example, in Sect. IV we have considered
the
case of two adjacent CuO planes in the high-T$_{c}$ superconductor
La$_{2-x}%
$Sr$_{x}$CuO$_{4}$. We have shown that entanglement can produce
significant
deviations from the standard Fermi-Dirac distribution in the normal state.
It
also leads to a shift of the chemical potential with the concentration
that is
very close to one of a superconducting distribution at finite temperature.
The
results are in reasonable agreement with experimental data, indicating
that a
more detailed study may be important in understanding the
pseudogap-related phenomena.

\section*{Appendix}

The Grassmann coherent states of Eq. (\ref{zeta}) can also be expressed by
the
action of a unitary operator on the vacuum as
\begin{equation}
\left\vert \tau\right\rangle =\exp\left[  \sum_{k=1}^{L}(\tau_{k}%
a_{k}^{\dagger}-a_{k}\tau_{k}^{\ast})\right]  \left\vert 0\right\rangle .
\label{tauket}%
\end{equation}
The relationship between the Grassmann variables $\tau_{k}$ and
$\zeta_{k}$
can be obtained by observing that for each $k,$ the operators
\begin{equation}
S_{+}=\frac{\tau_{k}}{|\tau_{k}|}a_{k}^{\dagger}~,\;S_{-}=a_{k}\frac{\tau
_{k}^{\ast}}{|\tau_{k}|}~,\;S_{0}=a_{k}^{\dagger}a_{k}-\frac{1}{2}\,,
\end{equation}
with $|\tau_{k}|=\sqrt{\tau_{k}^{\ast}\cdot\tau_{k}}$ generate an $su(2)$
algebra, and that
\begin{align}
\exp\left(  \tau_{k}a_{k}^{\dagger}-a_{k}\tau_{k}^{\ast}\right)
\left\vert
0\right\rangle  &  =\exp\left[  \left\vert \tau_{k}\right\vert \left(
S_{+}-S_{-}\right)  \right]  \left\vert 0\right\rangle \\
&  =\cos\left\vert \tau_{k}\right\vert \exp\left(  \frac{\tan|\tau_{k}|}%
{|\tau_{k}|}\tau_{k}a_{k}^{\dagger}\right)  \left\vert 0\right\rangle
.\nonumber
\end{align}
Thus, the expression (\ref{tauket}) has the form of the coherent state
(\ref{zeta}) with $\zeta_{k}=\tau_{k}|\tau_{k}|^{-1}\tan|\tau_{k}|$.

With respect to the real parametrization introduced in Sect. II, the
Grassmann
coherent states provide a resolution of the identity operator
$\mathcal{I}$ of
the many-particle Hilbert space such that $\mathcal{I}=c_{P}\mathcal{P}$
with
\begin{equation}
\mathcal{P}=\int d\Omega\left|  \zeta\right\rangle \left\langle
\zeta\right|
\,,
\end{equation}
where $c_{P}$ is the normalization constant and $d\Omega$ is a volume
element
on the manifold of Grassmann coherent states.

If the model space contains only one fermion state $(L=1)$, the Fock space
consists of the vacuum $\left\vert 0\right\rangle $ and a one-particle
state
$\left\vert 1\right\rangle =a_{1}^{\dagger}\left\vert 0\right\rangle ,$
and
$\left\vert \zeta\right\rangle =\sqrt{\xi_{1}}\left(  \left\vert
0\right\rangle -\zeta_{1}\left\vert 1\right\rangle \right)
=\sqrt{\xi_{1}%
}\left(  \left\vert 0\right\rangle +\left\vert 1\right\rangle \zeta
_{1}\right)  $ with $\zeta_{1}=z_{1}w_{1}$, $z_{1}=\rho_{1}e^{i\phi_{1}}$,
and
$\xi_{1}=1/(1+\rho_{1}^{2}).$ In this case
\begin{equation}
\left\langle n_{1}|\zeta\right\rangle =\sqrt{\xi_{1}}\zeta_{1}^{n_{1}},
\end{equation}
where $n_{1}=0,1$. The variables $\xi_{1}$, $\phi_{1}$ are related to
$p_{1}$
and $q_{1}$ defined in Eq. (\ref{pkqk}) by a canonical transformation, and
if
the volume element is chosen as
$d\Omega=dp_{1}dq_{1}/\hbar=d\xi_{1}d\phi_{1}%
$, then the off-diagonal terms are eliminated by the integration over
$\phi_{1},$ leaving
\begin{equation}
\left\langle i\right\vert \mathcal{P}\left\vert k\right\rangle =2\pi
\delta_{ik}\int d\xi_{1}\xi_{1}(\delta_{i0}+\rho_{1}^{2}w_{1}w_{1}^{\ast
}\delta_{i1})
\end{equation}
Because $\rho_{1}^{2}=\xi_{1}^{-1}-1$ and
\begin{equation}
\int_{0}^{1}d\xi_{1}\xi_{1}=\int_{0}^{1}d\xi_{1}\xi_{1}(\frac{1}{\xi_{1}%
}-1)=\frac{1}{2}%
\end{equation}
then $\left\langle i\right\vert \mathcal{P}\left\vert k\right\rangle
=\pi\delta_{ik}$, which shows that $c_{P}=1/\pi$.

Similarly, it can be shown that in general, the overlap between the
Grassmann
coherent state of Eq. (\ref{zeta}) and a many-body state $\left\vert
\left[
n\right]  \right\rangle \equiv\left\vert n_{1}n_{2}...n_{L}\right\rangle
=(a_{1}^{\dagger})^{n_{1}}(a_{2}^{\dagger})^{n_{2}}...(a_{L}^{\dagger}%
)^{n_{L}}\left\vert 0\right\rangle $, $n_{1},n_{2},...n_{L}=0,1$, is, up
to a
$\pm1$ factor,%
\begin{equation}
\left\langle n_{1}n_{2}...n_{L}|\zeta\right\rangle =\prod_{k=1}^{L}\sqrt
{\xi_{k}}\zeta_{k}^{n_{k}}.
\end{equation}
By choosing $d^{L}\Omega=\prod_{k=1}^{L}d\xi_{k}d\phi_{k}$, we get
\begin{equation}
\left\langle n_{1}n_{2}...n_{L}\right\vert \int d^{L}\Omega\left\vert
\zeta\right\rangle \left\langle \zeta|m_{1}m_{2}...m_{l}\right\rangle =\pi
^{L}\delta_{\lbrack n][m]}~,
\end{equation}
which proves the decomposition of the identity
\begin{equation}
\mathcal{I}=\frac{1}{\pi^{L}}\int d^{L}\Omega\left\vert \zeta\right\rangle
\left\langle \zeta\right\vert ~.
\end{equation}
Using this formula we can express the trace of a many-body operator $A$ as
an
integral over Grassmann coherent states. Thus,
\begin{align}
\mathrm{Tr}(A) &  =\mathrm{Tr}(A\mathcal{I})=\Sigma_{\lbrack
n]}\left\langle
[n]\right\vert A\mathcal{I}\left\vert [n]\right\rangle \\
&  =\frac{1}{\pi^{L}}\int d^{L}\Omega\Sigma_{\lbrack n]}\left\langle
[n]\right\vert A\left\vert \zeta\right\rangle \left\langle \zeta\right.
\left\vert [n]\right\rangle .\nonumber
\end{align}
However, because the $\left\vert [n]\right\rangle $ constitute a complete
set
of many-particle configurations in the selected subspace, $\sum_{[n]}%
\left\langle \zeta|[n]\right\rangle \left\langle [n]\right\vert
=\left\langle
\zeta\right\vert $, and
\begin{equation}
\mathrm{Tr}(A)=\frac{1}{\pi^{L}}\int d^{L}\Omega\left\langle
\zeta\right\vert
A\left\vert \zeta\right\rangle .
\end{equation}

\begin{acknowledgments}
Support of this research by the Natural Sciences and Engineering Research
Council of Canada is gratefully acknowledged.
\end{acknowledgments}

\newpage\textbf{Figure captions} \\[.5cm]Fig. 1. Planar electron density
as a
function of energy. (a) normal Fermi-Dirac distribution; (b)
superconducting
at finite temperature; (c) entangled Fermi-Dirac. \newline Fig. 2.
Chemical
potential shift as a function of hole concentration. (*) experimental
values;
(a) normal Fermi-Dirac distribution; (b) superconducting at finite
temperature; (c) entangled Fermi-Dirac; (d) quadratic polynomial fit of
the
experimenal data.


\begin{thebibliography}{99}
%


\bibitem {kirk}J. G. Kirkwood, Phys. Rev. \textbf{44}, 31 (1933).

\bibitem {mar}J. L. Martin, Proc. Roy. Soc. (London) \textbf{A251}, 536,
543 (1959).

\bibitem {htc1}J. G. Bednorz and K. A. M\"uller, Z. Phys. \textbf{B64},
189 (1986).

\bibitem {tt1}J. A. Simmons, M. A. Blount, J. S. Moon, S. K. Lyo, W. E.
Baca,
J. R. Wendt, J. L. Reno and M. J. Hafich, J. Appl. Phys. \textbf{84} 5626
(1998).

\bibitem {tt2}S. K. Lyo, Phys. Rev. \textbf{B 61} 8316 (2000).

\bibitem {kramer}P. Kramer and M. Saraceno, \textit{Geometry of the
Time-Dependent Variational Principle in Quantum Mechanics}, Lecture Notes
in
Physics, vol. 140, Springer, New York, 1981.

\bibitem {schw}J. Schwinger, Phys. Rev. \textbf{92}, 1283 (1953).

\bibitem {klau}G. Junker and J. R. Klauder, quant-ph/9708027 - Eur. Phys.
J.
\textbf{C4}, 173 (1998).

\bibitem {mit}M. Imada, A. Fujimori and Y. Tokura, Rev. Mod. Phys.
\textbf{70}, 1039 (1998).

\bibitem {ber}F. A. Berezin, \textit{The Method of Second Quantization},
Academic Press, New York, (1966).

\bibitem {casa}R. Casalbuoni, Il Nuovo Cimento \textbf{33A}, 115, 389
(1976).

\bibitem {abe}S. Abe and S. Naka, Prog. Theor. Phys. \textbf{72}, 881
(1984).

\bibitem {doran}C. Doran, D. Hestens, F. Sommen and N. Van Acker, J. Math.
Phys. \textbf{34}, 3642 (1993).

\bibitem {bay}W. E. Baylis, \textit{Electrodynamics - A Modern Geometric
Approach}, Birkh\"{a}user, Boston, (1999), p. 8.

\bibitem {loun}P. Lounesto, \textit{Clifford Algebras and Spinors, 2e,
}Cambridge University Press, Cambridge, UK, (2001).

\bibitem {somm}A. Sommerfeld, \textit{Thermodynamik und Statistik}, 3
Aufl.,
Akad. Verl. Ges., Leipzig, (1965).

\bibitem {zh}L. Zheng, M. W. Ortalano, and S. Das Sarma, Phys. Rev.
\textbf{B55}, 4506 (1997).

\bibitem {kastner}M. A. Kastner, R. J. Birgeneau, G. Shirane and Y. Endoh,
Rev. Mod. Phys. \textbf{70}, 897 (1998).

\bibitem {kim}W. Kim and J. P. Carbotte, cond-mat/0010324

\bibitem {pav}E. Pavarini, I. Dasgupta, T. Saha-Dasgupta, O. Jepsen and O.
K.
Andersen, cond-mat/0012051

\bibitem {bs}T. Timusk and B. W. Statt, Rep. Prog. Phys. \textbf{62}, 61
(1999).

\bibitem {pwa}P. W. Anderson, cond-mat/0108522

\bibitem {ck}S. Chakravarty, A. Sudbo, P. W. Anderson and S. Strong,
Science
\textbf{261}, 337 (1993).

\bibitem {cooper}Y. Zha, S. L. Cooper and D. Pines, Phys. Rev. \textbf{B
53},
8253 (1996).

\bibitem {loram}J. W. Loram, K. A. Mirza, J. R. Cooper and J. L. Tallon,
Physica C \textbf{282}, 1405 (1997).

\bibitem {fuji}A. Fujimori, A. Ino, T. Yoshida, T. Mizokawa, Z.-X. Shen,
C.
Kim, T. Kakeshita, H. Eisaki and S. Uchida, cond-mat/0011293

\bibitem {blatt}J. M. Blatt, \textit{Theory of Superconductivity},
Academic
Press, New York, (1964), p. 237.

\bibitem {ino}A. Ino, T. Mizokawa, A. Fujimori, K. Tamasaku, H. Eisaki,
S. Uchida, T. Kimura T. Sasagawa, and K. Kishio, Phys. Rev. Lett.
\textbf{79}, 2101 (1997).

\bibitem {tfd}Y. Takahashi and H. Umezawa, Coll. Phen. \textbf{2}, 55
(1975).
\end{thebibliography}
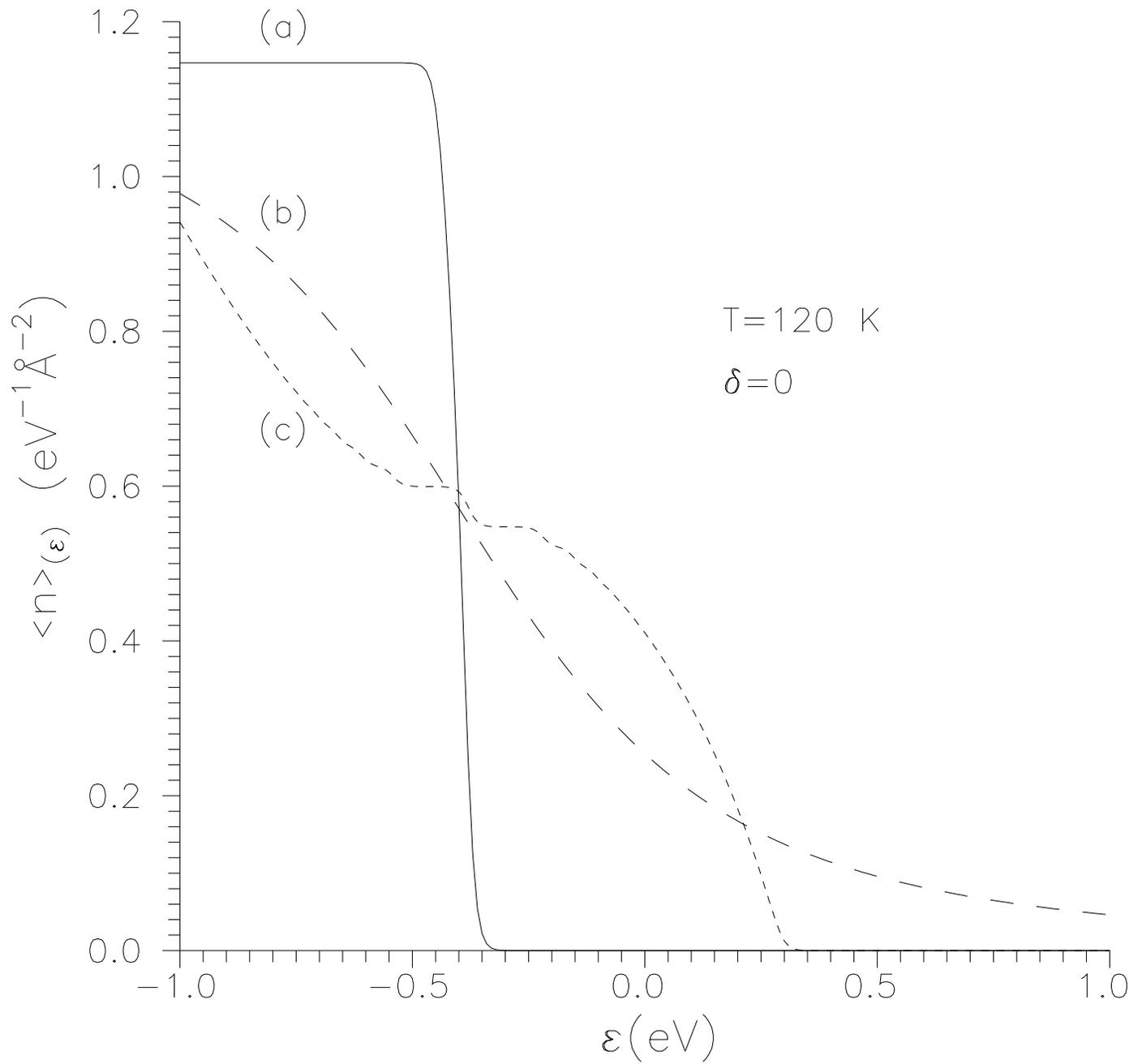
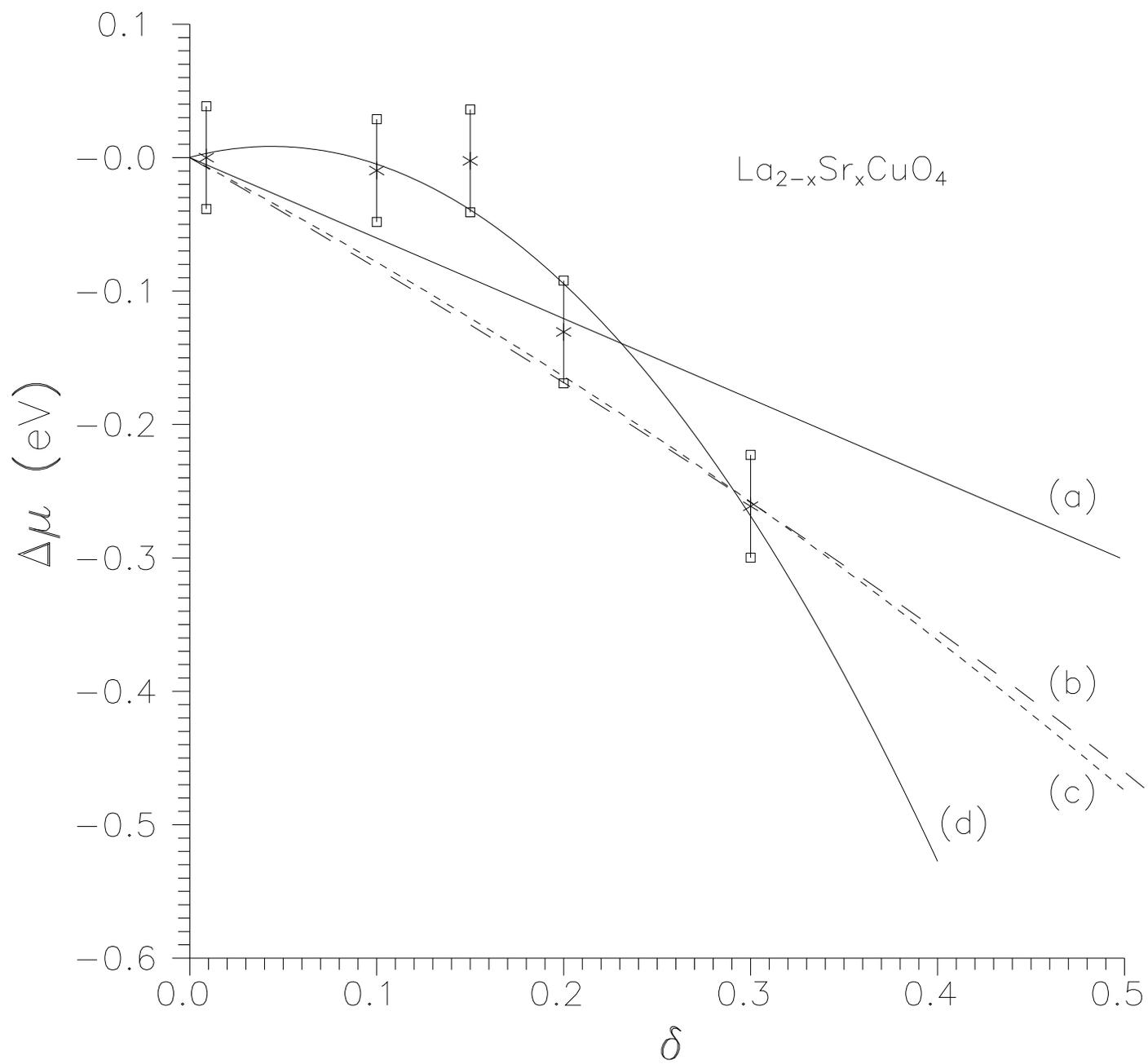
\end{document}